# Spatial Effects and Convergence Theory in the Portuguese Situation

Vítor João Pereira Domingues Martinho[1]

[1]Escola Superior Agrária, Instituto Politécnico de Viseu, Quinta da Alagoa,
Estrada de Nelas, Ranhados, 3500 - 606 VISEU
Centro de Estudos em Educação, Tecnologias e Saúde (CI&DETS)
**Portugal**
**e-mail:** vdmartinho@esav.ipv.pt

**Abstract**

This study analyses, through cross-section estimation methods, the influence of spatial effects and human capital in the conditional productivity convergence (product per worker) in the economic sectors of NUTs III of mainland Portugal between 1995 and 2002. To analyse the data, Moran's I statistics is considered, and it is stated that productivity is subject to positive spatial autocorrelation (productivity develops in a similar manner to productivity in neighbouring regions), above all, in agriculture and services. Industry and the total of all sectors present indications that they are subject to positive spatial autocorrelation in productivity. On the other hand, it is stated that the indications of convergence, specifically bearing in mind the concept of absolute $\sigma$ convergence, are greater in industry. Taking into account the estimation results, it is stated once again that the indications of convergence are greater in industry, and it can be seen that spatial spillover effects, spatial lag (capturing spatial autocorrelation through a spatially redundant dependent variable) and spatial error (capturing spatial autocorrelation through a spatially redundant error term), as well as human capital, condition the convergence of productivity in the various economic sectors of Portuguese region in the period under consideration (Martinho, 2011).

**Keywords:** Spatial Econometric; Growth Endogenous Theory; Portuguese Regions

**1. INTRODUCTION**

There are few known studies concerning conditional productivity convergence with spatial effects. Fingleton (2001), for example has found spatial correlation at the level of productivity when, using data from 178 regions of the European Union, he introduced spillover effects in a model of endogenous growth. Abreu et al. (2004) have investigated the spatial distribution of the rates of total productivity growth of factors using exploratory analyses of spatial data and other techniques of spatial econometrics. The sample consists of 73 countries and covers the period from 1960 to 2000. They have found significant spatial correlation in the rates of total factor productivity growth, indicating that high and low values tend to concentrate in space, forming the so-called "clusters". They have also found high indications of positive spatial autocorrelation at the level of the total factor productivity, which has increased throughout the period of 1960 to 2000. This result could indicate a tendency to clustering with time.

There is, on the other hand, a variety of studies analysing conditional product convergence with spatial effects. Armstrong (1995) has defended that the evidence of convergence across European countries as mentioned by Barro and Sala-i-Martin is due to the omission of spatial autocorrelation in their analysis and bias resulting from the selection of European regions. Following on, Sandberg (2004), for example, has examined the hypothesis of absolute and conditional convergence across Chinese provinces in the period from 1985 to 2000 and found indications that there had been absolute convergence during the periods of 1985 to 2000 and 1985 to 1990. He has also found evidence that conditional convergence had been seen in the sub-period of 1990 to 1995, with signs of spatial dependency across adjacent provinces. Arbia et al. (2004) have studied the convergence of gross domestic product per capita among 125 regions of 10 European countries from 1985 to 1995, considering the influence of spatial effects. They concluded that the consideration of spatial dependency considerably improved the rates of convergence. Lundberg (2004) has tested the hypothesis of conditional convergence with spatial effects between 1981 and 1990 and, in contrast to previous results, found no clear evidence favouring the hypothesis of conditional convergence. On the contrary, the results foresaw conditional divergence across municipalities located in the region of Stockholm throughout the period and for municipalities outside of the Stockholm region during the 1990s.

This study seeks to test conditional productivity convergence (using as a proxy the product per worker) for each of the economic sectors of regions (NUTs III) of mainland Portugal from 1995 to 2002, through techniques of cross-section spatial econometrics. To do so, this study is structured in six parts: after this introduction, there follows the second part where some theoretical considerations of spatial econometrics are presented; in the third part the models considered are explained; in the fourth part the data based on techniques of spatial econometrics developed to explore spatial data are analysed; in the fifth part the estimates drawn up are presented and in the sixth the main conclusions obtained after this research are highlighted.

## 2. THEORETICAL CONSIDERATIONS ABOUT CONVERGENCE WITH SPATIAL EFFECTS

The Neoclassical Theory of absolute (or unconditional) convergence states that poor countries or regions with low capital/work ratios have a greater marginal productivity of capital and can therefore grow more than richer countries or regions, given the same level of saving and investment. In this context, the tendency is for disparities to decrease over a period of time, since there is a tendency that factor costs will be lower in poorer regions, and, as a result, the opportunities for capital profit will be higher in these regions in comparison with richer regions. Therefore, less developed regions attract more investment and tend to grow quicker, thus getting closer to the leading regions. In the long term, the differences in profit and the rates of growth become equal across regions, since the existence of free trade and perfect mobility of input encourage convergence. Consequently, in this theory, convergence to a steady-state is the rule and divergence is a short term transitory phenomenon which reflects a period of adjustment. Technical progress is exogenous and is treated as a public asset, freely available to poor regions, thus facilitating the process of imitation and allowing for rapid growth without the costs of innovation. At an empirical level the Neoclassic approach to absolute convergence is based on the concept of $\beta$ convergence. The hypothesis of Neoclassic convergence is consistent with Solow's exogenous growth model (1956), where growth is determined by the exogenous offer of inputs, displaying constant profits or scaled decreases.

The concept of $\sigma$ convergence measures the dispersion of profit per capita or productivity across different economies over a period of time and the concept of $\beta$ convergence predicts the inverse relation between profit growth per capita or productivity and its initial level (through cross-section estimates). The evidence of $\sigma$ convergence is useful, since it allows periods of convergence or divergence over time to be observed. The existence of $\beta$ convergence is different, since it shows the rate of convergence across countries (regions), and implies that poor countries (regions) grow at a greater rate than rich countries (regions. The two measurements are complementary, but not exclusive. $\beta$ convergence is a necessary condition, but not sufficient in itself for there to be $\sigma$ convergence (Sala-i-Martin, 1996). To sum up, the concept of $\beta$ convergence is used more to predict absolute and conditional convergence. It should also be mentioned that the concept of $\beta$ convergence was first introduced by Barro and Sala-i-Martin (1991) to distinguish the concept of $\sigma$ convergence which measures, as previously stated, the dispersion of growth per capita using standard deviation or the coefficient of variation of a particular sample.

More recently, the concept of conditional convergence associated to the Theory of Endogenous Growth has been introduced. This emphasises the importance of human capital, innovation and increased profits as conditioning factors of convergence (Barro, 1991). Economies converge to different steady states which depend on the stock of human capital and the accumulation of physical capital, among others. This Theory predicts, therefore, a quicker growth for economies which have not reached their steady state. Empirical studies support that the hypothesis of absolute convergence is only seen in special cases where the sample involves economies with a high degree of homogeneity and across regions of the same country. This is known as the "convergence club" hypothesis (Chatterji, 1992). The majority of studies present results which support the hypothesis of conditional convergence, where, besides the level of profit per capita or initial productivity, the accumulation of physical and human capital and the innovation activities were the most significant conditioning factors. .

The studies which have sought to analyse conditional productivity convergence with spatial effects, have considered the base model as follows:

$$p = \rho W p + b \ln P_0 + \varepsilon \;^1, \text{conditional productivity convergence,} \quad (1)$$
$$\text{with spatial effects}$$

where p is rate of growth of sector productivity across various regions, $P_0$ is initial productivity, W is the matrix of distances, b is the convergence coefficient, $\rho$ is the autoregressive spatial coefficient (of the spatial lag component) and $\varepsilon$ is the error term (of the spatial error component, with, $\varepsilon = \lambda W \varepsilon + \xi$). The spatial lag and spatial error components are two spatial components which capture spatial effects in the redundant dependent variable and the error term respectively.

A potential source of errors of specification in spatial econometric models comes from spatial heterogeneity (Lundberg, 2004). There are typically two aspects related to spatial heterogeneity, structural instability and heteroskedasticity. Structural instability has to do with the fact that estimated parameters are not consistent across regions. Heteroskedasticity has to do with errors of specification which lead to non-constant variances in the error term. To prevent these types of errors of specification and to test for the existence of spatial lag and spatial error components in models, the results are generally complemented with specification tests. One of the tests is the Jarque-Bera test which tests the stability of parameters. The Breuch-Pagan and Koenker-Bassett, in turn, tests for heteroskedasticity. The second test is the most suitable when normality is rejected by the Jarque-Bera test. To find out if there are spatial lag and spatial error components in the models, two robust Lagrange Multiplier tests are used ($LM_E$ for "spatial error" and $LM_L$ for "spatial lag"). In brief, the $LM_E$ tests the null hypothesis of spatial non-correlation against the alternative of the spatial error model ("lag") and $LM_L$ tests the null hypothesis of spatial non-correlation against the alternative of the spatial lag model to be the correct specification.

---

[1] Starting from the coefficient of convergence $b = (1 - e^{-\beta T})$ it is possible to obtain the rate of convergence $\beta = -\ln(1-b)/T$.

According to the recommendations of Florax et al. (2003) and using the so-called strategy of classic specification, the procedure for estimating spatial effects should be carried out in six steps: 1) Estimate the initial model using the procedures using OLS; 2) Test the hypothesis of spatial non-dependency due to the omission spatially redundant variables or spatially autoregressive errors, using the robust tests $LM_E$ and $LM_L$; 3) If none of these tests has statistical significance, opt for the estimated OLS model, otherwise proceed to the next step, 4) If both tests are significant, opt for spatial lag or spatial error specifications, whose test has greater significance, otherwise go to step 5;; 5) If $LM_L$ is significant while $LM_E$ is not, use the spatial lag specification; 6) If $LM_E$ is significant while $LM_L$ is not, use the spatial error specification.

A test usually used to indicate the possibility of global spatial autocorrelation is the Moran's I test[2].
Moran's I statistics is defined as:

$$I = \frac{n}{S} \frac{\sum_i \sum_j w_{ij}(x_i - u)(x_j - u)}{\sum_i (x_i - u)^2} \text{, Moran's global autocorrelation test} \qquad (2)$$

where n is the number of observations and $x_i$ and $x_j$ are the observed rates of growth in the locations i and j (with the average u). S is the constant scale given by the sum of all the distances: $S = \sum_i \sum_j w_{ij}$.

When the normalisation of weighting on the lines of the matrix for distances is carried out, which is preferable, S equals n, since the weighting of each line added up should be equal to the unit, and the statistical test is compared with its theoretical average, $I=-1/(n-1)$. Then $I \to 0$, when $n \to \infty$. The null hypothesis $H_0$: $I=-1/(n-1)$ is tested against the alternative hypothesis $H_A$: $I \neq -1/(n-1)$. When $H_0$ is rejected and $I>-1/(n-1)$ the existence of positive spatial autocorrelation can be verified. That is to say, the high levels and low levels are more spatially concentrated (clustered) than would be expected purely by chance. If $H_0$ is rejected once again, but $I<-1/(n-1)$ this indicates negative spatial autocorrelation.

Moran's I local autocorrelation test investigates if the values coming from the global autocorrelation test are significant or not:

$$I_i = \frac{x_i}{\sum_i x_i^2} \sum_j w_{ij} x_j \text{, Moran's local autocorrelation test} \qquad (3)$$

where the variables signify the same as already referred to by Moran's I global autocorrelation test.

### 3. MODEL OF CONDITIONAL CONVERGENCE WITH SPATIAL EFFECTS

Bearing in mind the aforementioned theoretical considerations, what is presented next is the model used to analyse conditional productivity convergence with spatial effects and with human capital, at a sector and regional level in mainland Portugal:

$$(1/T)\log(P_{it}/P_{i0}) = \alpha + \rho W_{ij} p_{it} + b \log P_{i0} + X'\gamma + \varepsilon_{it} \text{, with } \alpha > 0 \text{ e } \beta < 0 \qquad (4)$$

In this equation (4) P is sector productivity, p is the rate of growth of sector productivity in various regions, W is the matrix of distances, X is the vector of variables which represent human capital (levels of schooling – primary, secondary and higher) b is the convergence coefficient, $\rho$ is the autoregressive spatial coefficient (of the spatial lag component) and $\varepsilon$ is the error term (of the spatial error component, with, $\varepsilon = \lambda W \varepsilon + \xi$). The indices i, j and t, represent the regions under study, the neighbouring regions and the period of time respectively.

### 4. DATA ANALYSIS

The data referring to gross growth value to base prices and employment were obtained in the Regional

---

[2] A similar, but less well-known test is Geary's C test (Sandberg, 2004).

Accounts of the National Statistics Institute. To carry out the cross-section estimations, the GeoDa[3] software was used. Also used were data concerning the level of schooling – primary, secondary or higher education, obtained from the 2001 Census of the National Statistics Institute.

What follows is an analysis of the data and $\sigma$ convergence, for the product per worker as proxy of the productivity of work in the period 1995 to 2002 in the various economic sectors of the regions (NUTs III) of mainland Portugal. The data analysis is carried out while considering, in the various economic sectors, the values of the productivity ratio of each of the regions under consideration, in relation to average productivity in mainland Portugal. The values of $\sigma$ convergence were calculated using the year on year variation coefficient across the different regions. It also seeks to identify the existence of spatial autocorrelation by using Moran Scatterplots for over all spatial autocorrelation and LISA Maps for local spatial autocorrelation.

### 4.1. $\sigma$ CONVERGENCE OF SECTOR PRODUCTIVITY ACROSS NUTS III

When considering the values base referring to the percentage values of productivity for each of the economic sectors in relation to the average of each of the NUTs III in mainland Portugal, the following should be noted:

Agriculture in the North presents the highest relative productivity in Greater Porto (with above average values) and the Douro (below, but close to, average value). On the other hand the lowest relative productivity values, with a tendency to decrease, are presented in Minho-Lima. In the Centre, Beira Interior Sul presents the highest productivity, which is above average, and Serra da Estrela shows the lowest values. Lisbon and Vale do Tejo to Lezíria do Tejo present the highest, and above average, productivity and the Médio Tejo shows the lowest rates, which have remained reasonably constant over time. The Alentejo and the Algarve present in all relative productivities, above average values and with a tendency to remain reasonably constant or to increase. Upon analysing the values of $\sigma$ convergence (Table and Graph 1) the tendency for divergence as previously described is confirmed.

As far as the industry analysed is concerned, it can be seen that there is a certain tendency for convergence of productivities in this sector across the different NUTs III of mainland Portugal, which is confirmed by the values of $\sigma$ convergence (Table and Graph 1).

As for services, and for the total of regional economies for each of the NUTs III, it can be seen from the values presented in Graph 3 (appended) that there was a tendency for a degree of convergence across the relative productivities in these economic sectors over the period of study, although the evidence for such convergence is slight, as can be confirmed by the values presented in Table and Graph 1 for $\sigma$ convergence.

**Table 1:** Values of $\sigma$ convergence by sectors of activity, across NUTs III
From 1995 to 2002

|  | 1995 | 1996 | 1997 | 1998 | 1999 | 2000 | 2001 | 2002 |
|---|---|---|---|---|---|---|---|---|
| **Agriculture** | 0,54 | 0,56 | 0,64 | 0,62 | 0,58 | 0,59 | 0,66 | 0,74 |
| **Industry** | 0,43 | 0,41 | 0,45 | 0,39 | 0,36 | 0,32 | 0,29 | 0,31 |
| **Services** | 0,10 | 0,10 | 0,10 | 0,12 | 0,12 | 0,11 | 0,10 | 0,10 |
| **Total** | 0,22 | 0,22 | 0,24 | 0,23 | 0,23 | 0,22 | 0,21 | 0,21 |

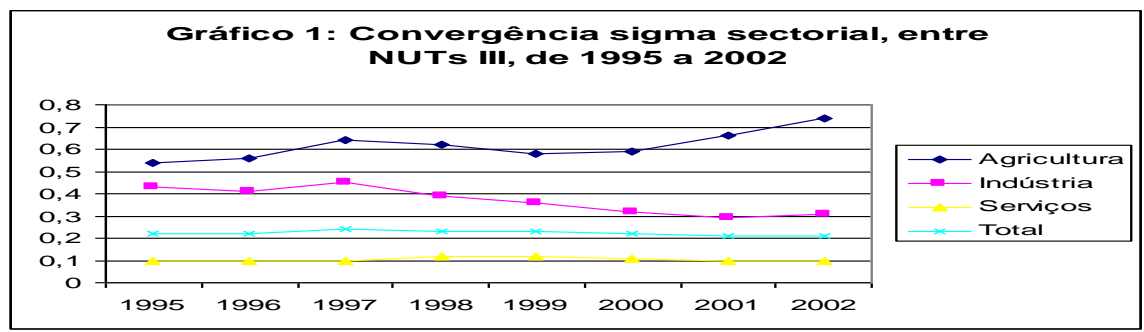

Gráfico 1: Convergência sigma sectorial, entre NUTs III, de 1995 a 2002

### 4.2. CROSS-SECTION DATA ANALYSIS

The four Scatterplots, (showing the relation between the growth of productivity and initial productivity for each of the sectors) presented below, allow for an analysis of productivity convergence for each of the economic sectors of the Portuguese NUTs III, with average values for the period 1995 to 2002.

---

[3] Available at http://geodacenter.asu.edu/

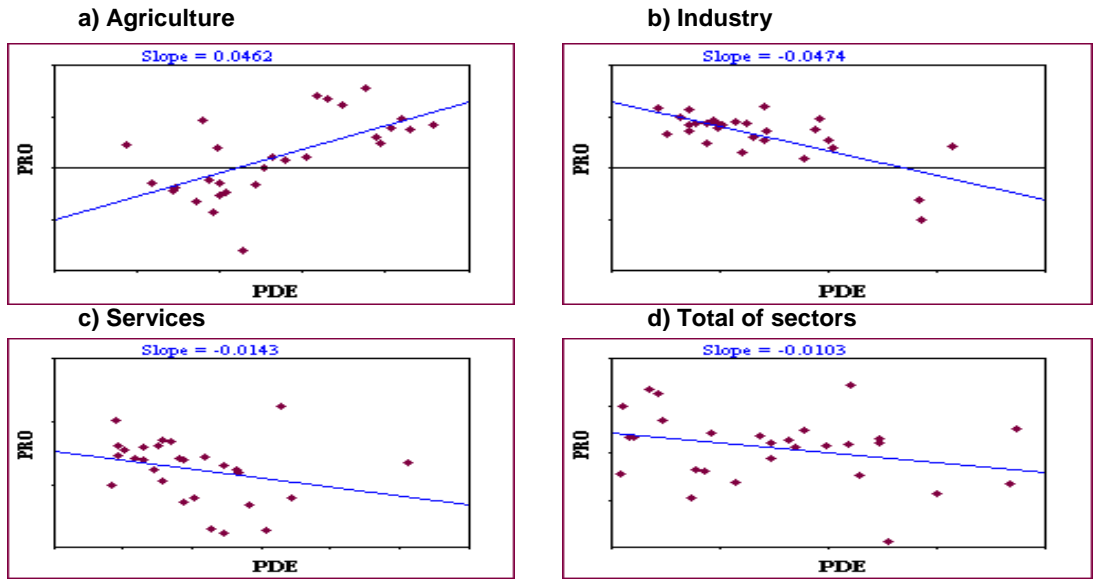

**Note:** PRO = Productivity;
PDE = Initial productivity.

***Figure 1:*** *Scatterplots of absolute convergence of productivity for each of the economic sectors (cross-section analysis, 28 regions)*

Analysing the four figures above confirms what has been previously shown, or, in other words, industry is the only economic sector which shows greater tendencies for absolute convergence.

The four Moran Scatterplots (showing the relationship between the dependent variable and the spatially redundant dependent variable) which are presented below, show Moran's I statistical values fro each of the economic sectors and for the total of sectors of the 28 NUTs for mainland Portugal from 1995 to 2002. The matrix $W_{ij}$ used is the matrix of the distances between the regions up to a maximum limit of 97 Km. This distance appeared to be the most appropriate to the reality of Portuguese NUTs III, given the signs of spatial autocorrelation encountered, (with an analysis of the data, bearing in mind namely Moran's I statistics, and with the estimation results carried out) in the analysis of robustness and behaviour of the various matrices of distance when considering alternative possibilities of maximum distances. For example, for agriculture and services which, as we shall see, are the sectors where the signs of autocorrelation are strongest, these indications cease to exist when the distances are significantly higher than 97 Km. On the other hand, the connectivity of the distance matrix is weaker for distances over 97 Km. Whatever the case, the choice of the best limiting distance to construct these matrices is always complex.

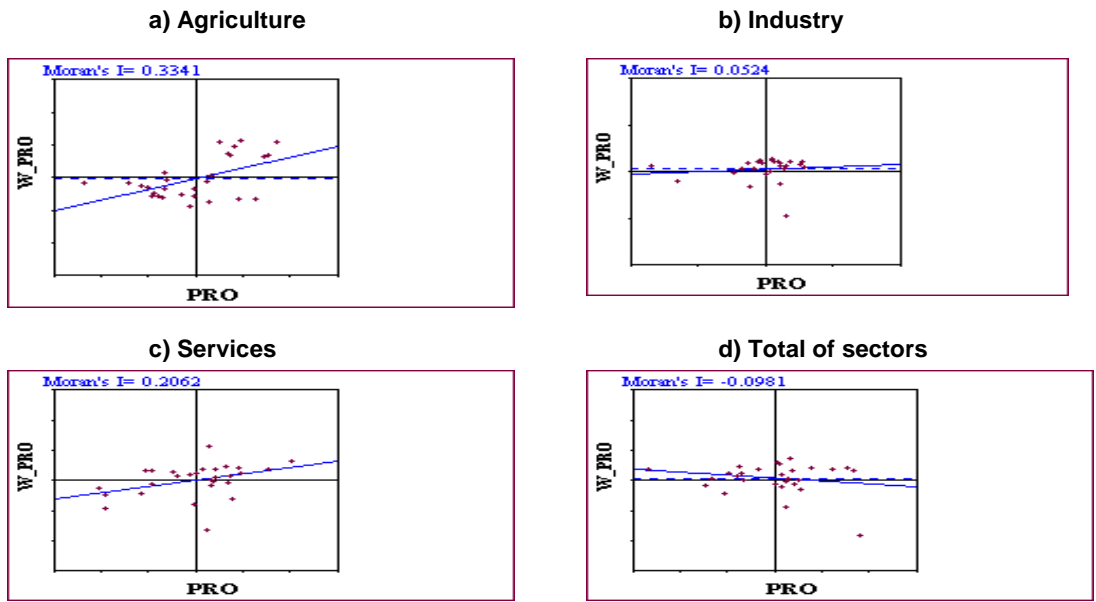

**Note:** W-PRO = Spatially redundant productivity;
PRO = Productivity.

***Figure 2:*** *"Moran Scatterplots" of productivity for each of the economic sectors (cross-section analysis, 28 regions)*

An analysis of the Moran Scatterplots shows that it is only in agriculture and services that the existence of global spatial autocorrelation can be seen in productivity and that there are few indications of the same occurring in industry, since Moran's I value is positive..

Figure 3 analyses the existence of local spatial autocorrelation with four LISA Maps, investigated under spatial autocorrelation and its significance locally (by NUTs III). The NUTs III with "high-high" and "low-low" values, correspond to the regions with positive spatial autocorrelation and with statistical significance, or, in other words, these are cluster regions where the high values ("high-high") or low values ("low-low") of two variables (dependent variable and redundant dependent variable) are spatially correlated given the existence of spillover effects. The regions with "high-low" and "low-high" values are "outliers" with negative spatial autocorrelation

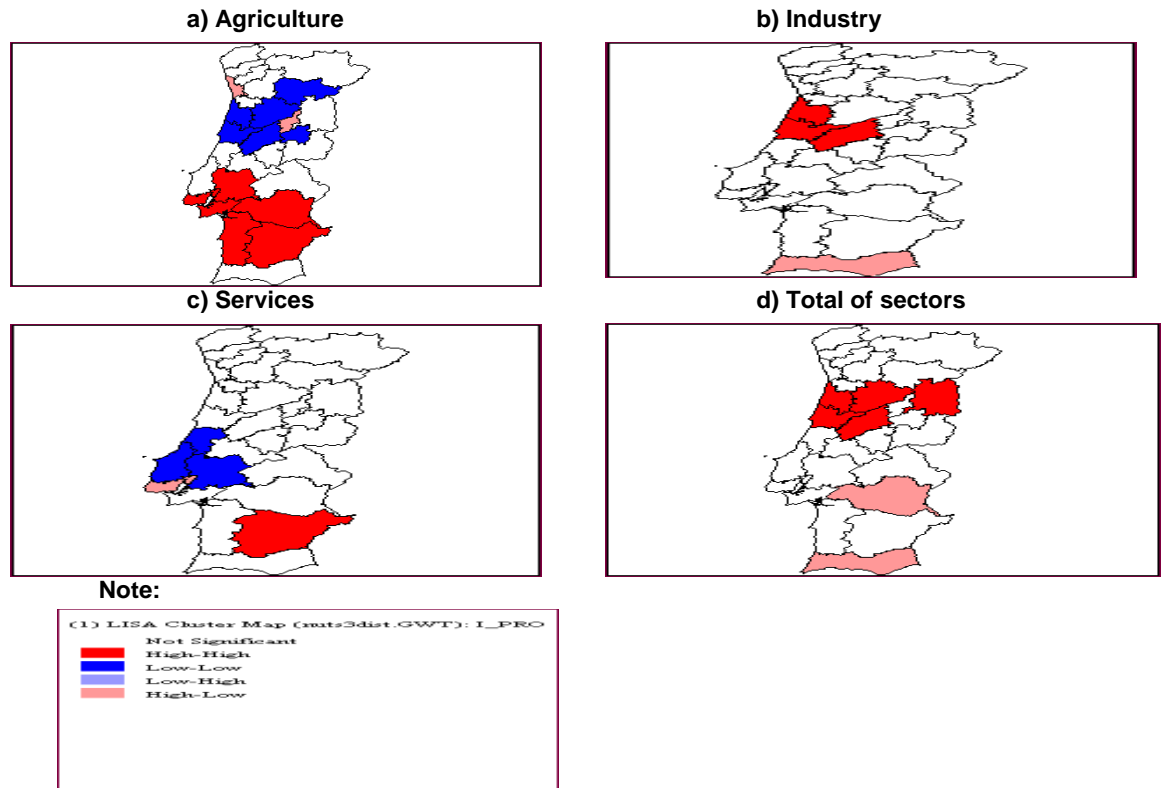

*Figure 3:* "LISA Cluster Map" of productivity for each of the economic sectors (cross-section analysis, 28 regions)

Analysing the LISA Cluster Maps above confirms what has been verified by the Moran Scatterplots, or, in other words, the indications of positive spatial autocorrelation are highest in agriculture and services. Agriculture shows signs of positive spatial correlation with high values in Greater Lisbon, around Greater Lisbon and the Alentejo and low values in the Centre-North region. Services present high values for the two variables in the Baixo Alentejo and low values in the region around Greater Lisbon. There are also some signs of positive spatial autocorrelation in these figures for industry and the total of sectors, more specifically with high values in some NUTs III of the Central region. In consideration of what has previously been referred to, spatial spillover effects in terms of productivity are non-existent in the North and the Algarve. This can be seen with high values in the Centre for industry and the total of sectors and with low values for agriculture. High values can be seen in Lisbon and Vale do Tejo for agriculture and low values for services. Positive spatial autocorrelation in the Alentejo can be seen with high values for agriculture and services. These signs of positive spatial autocorrelation as described for each of the economic sectors included in various NUTs III could be an indication of sector similarities in productive structure in each of the strips of land, given the example of the existence of spatial spillover effects for agriculture in the Alentejo.

**5. EMPIRICAL EVIDENCE FOR CONDITIONAL CONVERGENCE OF PRODUCTIVITY, CONSIDERING THE POSSIBILITY THAT THERE ARE SPATIAL EFFECTS**

What follows is the presentation of empirical evidence of the existence of conditional productivity convergence for each of the economic sectors of the Portuguese NUTs III from 1995 to 2002, based on cross-section estimates. The cross-section estimates were carried out using the Least Square (OLS) method and the Maximum Likelihood (ML) method. The use of the Maximum Likelihood method is based on the fact that the associated authors advise using it when there is the presence of spatial spillover effects.

### 5.1. EMPIRICAL EVIDENCE WITH SPATIAL EFFECTS AS CONDITIONING VARIABLES

This part of the study will examine the procedures of specification by Florax et al. (2003) and will firstly examine through OLS estimates, the relevance of proceeding with estimate models with spatial lag and spatial error components with recourse to LM specification tests.

The results concerning OLS estimates of conditional convergence with tests of spatial specification are present in Table 2, which follows.

**Table 2:** OLS estimation results for the equation of absolute convergence with spatial specification tests

$$(1/T)\log(P_{it}/P_{i0}) = \alpha + b\log P_{i0} + \varepsilon_{it}$$

|  | Con. | Coef. b | JB | BP | KB | M'I | $LM_l$ | $LMR_l$ | $LM_e$ | $LMR_e$ | $\bar{R}^2$ | N.O. |
|---|---|---|---|---|---|---|---|---|---|---|---|---|
| **Agriculture** | -0.399* (-3.974) | 0.046* (4.082) | 0.234 | 1.248 | 0.926 | -0.078 | 0.343 | 3.679** | 0.492 | 3.827** | 0.367 | 28 |
| **Industry** | 0.490* (5.431) | -0.047* (-5.090) | 0.971 | 17.573* | 13.065* | 0.120** | 0.003 | 0.863 | 1.149 | 2.009 | 0.480 | 28 |
| **Services** | 0.181** (1.928) | -0.014 (-1.479) | 0.031 | 4.627* | 4.094* | 0.092 | 1.499 | 4.924* | 0.673 | 4.098* | 0.042 | 28 |
| **Total of sectors** | 0.138* (2.212) | -0.010 (-1.559) | 0.437 | 0.296 | 0.271 | -0.141 | 2.043 | 0.629 | 1.593 | 0.180 | 0.050 | 28 |

Note: JB, Jarque-Bera test; BP, Breusch-Pagan test; KB, Koenker-Bassett test: M'I, Moran's I; $LM_l$, LM test for spatial lag component; $LMR_l$, robust LM test for spatial lag component; $LM_e$, LM test for spatial error component; $LMR_e$, robust LM test for spatial error component; $R^2$, coefficient of adjusted determination; N.O., number of observations; *, statistically significant to 5%; **, statistically significant to 10%.

This conforms to what has been previously seen in the data analysis, or, in other words, productivity convergence is only seen in industry, although the values of the convergence coefficient present indications of heteroskedasticity, according to the BP and KB tests. Agriculture presents clear signs of divergence, since the convergence coefficient is positive and statistically significant. Convergence in the productivity sector will be conditioned by spillover and spatial error effects in agriculture eventually and spill over and spatial lag effects in services, according to the LM tests.

Table 3 presents the results of the estimates of spillover and spatial error effects for agriculture and spillover and spatial lag effects for services.

**Table 3:** ML estimation results for the equation of conditional convergence to spatial effects

$$(1/T)\log(P_{it}/P_{i0}) = \alpha + \rho W_{ij} p_{it} + b\log P_{i0} + \varepsilon_{it}$$

|  | Constant | Coefficient | Spatial coefficient | Breusch-Pagan | $\bar{R}^2$ | N.Observations |
|---|---|---|---|---|---|---|
| **Agriculture** | -0.460* (-6.419) | 0.053* (6.558) | -0.496 (-1.405) | 0.915 | 0.436 | 28 |
| **Services** | 0.122 (1.365) | -0.010 (-1.065) | 0.327 (1.268) | 4.884* | 0.138 | 28 |

Note: *, statistically significant to 5%; **, statistically significant to 10%; ***, spatial coefficient of the spatial error model for agriculture and spatial lag model for services.

The convergence coefficient for agriculture is similar to what is presented in Table 2, although it has improved slightly statistically. In services the convergence coefficient is slightly different in terms of values obtained and statistical significance. On the other hand, the coefficients of spatial variables have no statistical significance. As a result, convergence in agriculture and services is not conditioned by spatial effects.
.
### 5.2. EMPIRICAL EVIDENCE WITH HUMAN CAPITAL AS A STRUCTURAL VARIABLE

Table 4 presents a series of estimates for conditional sector productivity convergence, with the level of schooling as a proxy for human capital. Three levels of schooling were considered (primary, secondary and higher education) represented by different variables. These variables were obtained through the percentage of the population with each level of schooling in relation to the total number of people, taking into account the data from the Census 2001. Different estimates for each sector were carried out for level of schooling so as to avoid problems of multicollinearity.

**Table 4:** Empirical evidence of the importance of the level of schooling in the convergence of productivity in the various economic sectors

$$(1/T)\log(P_{it}/P_{i0}) = \alpha + b\log P_{i0} + X'\gamma + \varepsilon_{it}$$

|       | Con.              | Coef.1              | Coef.2              | JB    | BP       | KB       | M'I     | LM$_l$  | LMR$_l$ | LM$_e$ | LMR$_e$ | R$^2$ | N.O. |
|-------|-------------------|---------------------|---------------------|-------|----------|----------|---------|---------|---------|--------|---------|-------|------|
| **Agriculture** | | | | | | | | | | | | | |
| Prim. | -0.200<br>(-1.552) | 0.037*<br>(3.302)   | -0.220*<br>(-2.249) | 8.486* | 5.007**  | 2.054    | -0.089  | 0.243   | 3.284** | 0.632  | 3.672** | 0.453 | 28 |
| Sec.  | -0.440*<br>(-4.401)| 0.040*<br>(3.508)   | 0.253<br>(1.684)    | 8.890* | 7.908*   | 3.232    | -0.112  | 0.129   | 3.723** | 0.996  | 4.591*  | 0.409 | 28 |
| High..| -0.370*<br>(-3.882)| 0.039*<br>(3.477)   | 0.414*<br>(2.098)   | 1.085  | 2.466    | 1.526    | -0.053  | 0.672   | 3.914*  | 0.223  | 3.466** | 0.440 | 28 |
| **Industry** | | | | | | | | | | | | | |
| Prim. | 0.578*<br>(6.197) | -0.050*<br>(-5.700) | -0.116*<br>(-2.198) | 0.565  | 18.144*  | 12.359*  | 0.076   | 0.010   | 0.180   | 0.461  | 0.630   | 0.547 | 28 |
| Sec.  | 0.448*<br>(4.809) | -0.048*<br>(-5.212) | 0.118<br>(1.426)    | 0.746  | 13.761*  | 10.875*  | 0.109** | 0.049   | 0.339   | 0.943  | 1.234   | 0.500 | 28 |
| High..| 0.521*<br>(6.285) | -0.053*<br>(-6.062) | 0.271*<br>(2.544)   | 3.450  | 33.593*  | 16.957*  | 0.016   | 0.054   | 0.161   | 0.021  | 0.128   | 0.570 | 28 |
| **Services** | | | | | | | | | | | | | |
| Prim. | 0.371*<br>(2.059) | -0.032**<br>(-1.853)| -0.034<br>(-1.231)  | 0.323  | 6.990*   | 5.055**  | 0.101   | 1.890   | 6.694*  | 0.819  | 5.623*  | 0.061 | 28 |
| Sec.  | 0.234**<br>(1.801)| -0.021<br>(-1.435)  | 0.021<br>(0.596)    | 0.033  | 5.873**  | 5.031**  | 0.093   | 1.607   | 7.047*  | 0.685  | 6.125*  | 0.018 | 28 |
| High. | 0.284*<br>(2.203) | -0.025**<br>(-1.872)| 0.051<br>(1.157)    | 0.553  | 10.749*  | 7.736*   | 0.105   | 1.791   | 3.734** | 0.875  | 2.818** | 0.054 | 28 |
| **Total of sectors** | | | | | | | | | | | | | |
| Prim. | 0.307*<br>(3.405) | -0.024*<br>(-2.900) | -0.070*<br>(-2.427) | 0.662  | 0.302    | 0.402    | -0.078  | 2.239   | 2.672   | 0.482  | 0.914   | 0.201 | 28 |
| Sec.  | 0.188*<br>(2.816) | -0.018*<br>(-2.326) | 0.072**<br>(1.727)  | 0.775  | 0.223    | 0.290    | -0.075  | 1.572   | 1.952   | 0.448  | 0.828   | 0.118 | 28 |
| High..| 0.213*<br>(3.001) | -0.019*<br>(-2.461) | 0.106**<br>(1.929)  | 0.130  | 1.134    | 1.072    | -0.165  | 3.354** | 1.331   | 2.178  | 0.156   | 0.140 | 28 |

Note: Prim., estimate with primary education; Sec., estimate with secondary education; High., estimate with higher education; Con., constant; Coef.1, coefficient of convergence; Coef. 2 coefficient of level of schooling; JB, Jarque-Bera test; BP, Breusch-Pagan test; KB, Koenker-Bassett test: M'I, Moran's I; LM$_l$, LM test for spatial lag component"; LMR$_l$, robust LM test for spatial lag component; LM$_e$, LM test for spatial error component; LMR$_e$, robust LM test for spatial error component; R$^2$, r squared adjusted; N.O., number of observations *, statistically significant to 5%; **, statistically significant to 10%.

In agriculture, for the three levels of schooling, the indications of divergence seen in the results presented in Table 2 are maintained, since the coefficients for convergence present a positive sign with statistical significance, although the values are slightly lower, which is a sign that the level of schooling productivity convergence in this sector, albeit slightly. On the other hand, as could be expected, primary education has a negative effect on the growth of productivity in agriculture for the period 1995 to 2002, while higher education has a positive effect. Therefore, the progress in the level of schooling in this sector improves productivity performances. As far as the LM test of specification are concerned, with the exception of the results obtained from the estimations of higher education, all figures confirm the previous results for this sector, or, in other words, the better specification of the model is with the spatial error component.

Industry confirms in these estimations the signs of productivity convergence across the NUTs III of mainland Portugal from 1995 to 2002, a fact which is only favoured by higher education (since the effect of higher education is positive and increases convergence). The non-existence of indications of spatial autocorrelation was also confirmed, given the values of the LM tests.

Contrary to what was seen in the results for absolute convergence, in these estimations of conditional productivity convergence in services, has the level of schooling as a conditioning variable. Some indications of convergence can be seen in the equations of primary education and higher education, which is sign that eventually convergence will be conditioned to human capital in this sector (since none of the coefficients associated to the variables of the level of schooling has statistical significance). On the other hand, taking into account the LM tests, it is confirmed that the better specification of the model is with the spatial lag component.

In the total of sectors, something similar to what was verified in services can see, or, in other words, the convergence coefficient has no statistical significance in the estimations for absolute convergence, but is present in the estimations for conditional productivity convergence with human capital. The difference is that here the coefficients of conditioning variables demonstrate statistical significance, an indicator that convergence in the total of sector sis conditioned by level of schooling.

Finally, it should be noted that the greatest marginal effect is through higher education schooling, which indicates that the higher the level of schooling, the greater the growth in productivity. .

### 6. CONCLUSIONS

This study has sought to test the convergence of productivity for each of the economic sectors (agriculture, industry, services and the totality of services) across the 28 regions (NUTs III) of mainland Portugal in the period of 1995 to 1999, with spillover, spatial lag and spatial error effects. To do so, data analysis and cross-section estimates (with average temporal values) have been carried out with the OLS and MI estimation methods, following the specification procedures indicated by Florax et al. (2003) who suggest that models are first tested with the OLS method, to test which is the better specification (spatial lag or spatial error) and then the spatial lag or spatial error is estimated with the ML method.

Considering the analysis of the cross-section data previously carried out, it can be seen that productivity (product per worker) is subject to positive spatial autocorrelation in agriculture and services (with Greater Lisbon, curiously, showing the greatest spatial spillover effects in agriculture than in services). Industry and the total of all sectors also show some signs of spatial autocorrelation. Also of note is the fact that the region surrounding Lisbon and the Alentejo will clearly have a great influence in the development of the economy with agriculture. On the other hand, it can be stated that the tendency for absolute productivity convergence is greatest in industry.

As far as cross-section estimates are concerned, it can be seen that sector by sector the tendency for productivity convergence is greatest in industry With reference to spatial autocorrelation it is also confirmed that this possibly exists in agriculture and services, when taking into account the LM tests. Following the procedures of Florax et al. (2003) the equation is estimated with the spatial error component for agriculture and the spatial lag component for services, and it can be seen that the consideration of these spatial effects doe snot significantly alter the results obtained previously with the OLS estimation.

The level of schooling as proxy for human capital conditioning productivity convergence, improves the value and statistical significance of convergence coefficients. On the other hand, above all the variable which represents higher education shows indications which directly favour the growth of productivity, since the coefficient associated to it presents in all economic sectors the greatest marginal positive effect.